\documentclass[10pt,fleqn]{article}
\usepackage{amsmath,graphicx,latexsym}

\newcommand{\ZZ}{\mbox{\sf Z\zr{-0.45}Z}}
\newcommand{\zr}[1]{\mbox{\hspace{#1em}}}

\begin{document}

\title{Real Jacobian Elliptic Function Parametrizations for a Genuinely Asymmetric
Biquadratic Curve}

\author{Apostolos Iatrou \thanks{{\copyright\mbox{ }Apostolos Iatrou}}
\thanks{Part of this paper is taken from the author's PhD thesis (La Trobe University).
The part taken was written under the supervision of K. A. Seaton.}
\thanks{{\sf email:
A.Iatrou@latrobe.edu.au\mbox{ }\mbox{ }apostolosiatrou@hotmail.com
}}}

\date{}

\maketitle

\begin{abstract}
\noindent We give real Jacobian elliptic function parametrizations
for a genuinely asymmetric biquadratic curve where the variables
and parameters are real.
\end{abstract}

\section{Introduction}

In a recent paper \cite{aijrii} a method was developed to simplify
any real elliptic asymmetric biquadratic curve\footnote{Following
\cite{aijrii}, we classify the biquadratic as {\em rational} if
at least one of the square root signs disappears when $y$ is
solved as a function of $x$ or $x$ is solved as a function of $y$,
meaning $y$ is a rational function of $x$ or $x$ is a rational
function of $y$. On the other hand, if neither square root sign
disappears we call the biquadratic {\em elliptic}. We call a
biquadratic symmetric when it is symmetric in the variables,
otherwise we call it asymmetric.}, i.e.
\begin{equation} \label{biquad}
    \bar{\alpha} X^{2}Y^{2} + \bar{\beta} X^{2}Y + \bar{\delta} XY^{2} + \bar{\gamma} X^{2}
    + \bar{\kappa} Y^{2} + \bar{\epsilon} XY + \bar{\xi} X + \bar{\lambda} Y + \bar{\mu} \,=\,0,
\end{equation}
to a canonical curve of the following form
\begin{equation} \label{normsym}
      x^{2}y^{2}+\gamma(x^{2}+y^{2})+\epsilon x y\pm1=0
\end{equation}
or
\begin{equation} \label{normasym}
      x^{2}y^{2}+\gamma(x^{2}-y^{2})+\epsilon xy-1=0,
\end{equation}
using a real invertible transformation $T (X,Y) : (X,Y) \mapsto
(x,y)=(f(X),g(Y))$ with $f$ and $g$ modular, and, correspondingly,
to simplify the dynamics on it from being generated by
\begin{equation} \label{1ma}
        X'=-X-\frac{\bar{\delta} Y^{2}+\bar{\epsilon} Y+\bar{\xi}} {\bar{\alpha} Y^{2}+\bar{\beta}
        Y+\bar{\gamma}}\, , \; \; Y'=-Y-\frac{\beta
        X'^{2}+\bar{\epsilon} X'+\bar{\lambda}} {\bar{\alpha} X'^{2}+\bar{\delta} X'+\bar{\kappa}}
\end{equation}
to
\begin{equation} \label{sym}
    x'=y\, , \; \;
    y'=-x-\frac{\epsilon y}{ y^2+\gamma}
\end{equation}
or
\begin{equation} \label{bmapasym}
    x'=-x-\frac{\epsilon y }{ y^2+\gamma}\, , \; \;
    y'=-y-\frac{\epsilon x' }{ {x'}^2-\gamma},
\end{equation}
respectively.

Futhermore, a procedure was outlined which one could use to
parametrize the {\em symmetric} canonical biquadratic curves given
in equation (\ref{normsym}), and consequently many of the
asymmetric biquadratic curves (\ref{biquad}), using Jacobian
elliptic functions\footnote{We refer the reader throughout this
section to e.g. \cite[pp. 18-31,284-285]{Byrd} for definitions
and properties of Jacobian elliptic functions.}. The results for
the parametrization when the variables and parameters are real
were given in \cite[Table 3]{aijrii}. However, two cases were not
parametrized. The fundamentally {\em asymmetric} canonical curve
given in equation (\ref{normasym}) was also not parametrized.  As
noted in \cite[Section 5]{aijrii}, however, if one allows complex
variables and parameters then all elliptic biquadratics can be
reduced to the canonical form given in equation (\ref{normsym})
(with $+$ sign) \footnote{This complex normal form for
biquadratics was also found independently in \cite{rcgo}.}. This
then can be parametrized by the Jacobian elliptic function
$\mathrm{sn}$, see \cite[pp. 471--473]{Bax}.

In this paper we give parametrizations for the fundamentally
asymmetric normal form for many asymmetric biquadratics given in
equation (\ref{normasym}) using Jacobian elliptic functions when
the variables and parameters are real. We also discuss the two
outstanding cases of the symmetric canonical curve given in
equation (\ref{normsym}) (with $+$ sign).

\section{Asymmetric Case}

Consider the asymmetric biquadratic given in equation
(\ref{normasym}) after making the replacement $\epsilon
\rightarrow 2\epsilon$, that is
\begin{equation}\label{basym}
  B(x,y)=x^2 y^2 +\gamma (x^2-y^2) +2\epsilon x y -1=0.
\end{equation}
Writing $y$ as a function of $x$, we get
\begin{equation} \label{byasafx}
    y = \frac{- \epsilon x \pm \sqrt{- \gamma x^4 +
    (\gamma^2 + \epsilon^2 + 1) x^2 - \gamma}}{x^2 - \gamma},
\end{equation}
while writing $x$ as a function of $y$, we get
\begin{equation} \label{bxasafy}
    x = \frac{- \epsilon y \pm \sqrt{\gamma y^4 +
    (\gamma^2 + \epsilon^2 + 1) y^2 + \gamma}}{y^2 + \gamma}.
\end{equation}
The nature of the curves given by (\ref{basym}), e.g. whether the
range of $x$ and $y$ is bounded or unbounded, can be determined by
looking at the quartics
\begin{equation} \label{bdescx}
    \Delta_x(x) = -\gamma x^4 + (\gamma^2 + \epsilon^2 + 1) x^2 - \gamma
\end{equation}
and
\begin{equation} \label{bdescy}
    \Delta_y(y) = \gamma y^4 + (\gamma^2 + \epsilon^2 + 1) y^2 + \gamma.
\end{equation}
The possibilities are listed in Figure 1, where we introduce
\begin{equation} \label{bdefellb}
{\cal B}_x = \frac{(\gamma^2+\epsilon^2 + 1)}{\gamma}
\end{equation}
and ${\cal B}_y=-{\cal B}_x$. Note that for $\gamma
> 0$ we get
\begin{equation}
  \gamma + \frac{1 }{ \gamma} \ge 2 \Rightarrow \gamma + \frac{1 }{ \gamma}
  + \frac{\epsilon^2 }{ \gamma} = \frac{\gamma^2+\epsilon^2+1 }{ \gamma}\ge
  2.
\end{equation}
Similarly, for $\gamma < 0$ we get
\begin{equation}
  \frac{\gamma^2+\epsilon^2+1 }{ \gamma} \le -2.
\end{equation}
Also note that $\Delta_x(x)$ $(\Delta_y(y))$ has three (one)
extrema if ${\cal B}_x \ge 2$  $({\cal B}_y \le -2)$ and one
(three) if ${\cal B}_x \leq -2$ $({\cal B}_y \ge 2)$.
\begin{figure}\label{52}
\centering
\begin{tabular}{|c||c|c|c|c|c|c|c|c|} \hline
Case & $\gamma$ & ${\cal B}_x$ & ${\cal B}_y$ & $\Delta_x(0)$ &  $\Delta_y(0)$ & $\Delta_x(x)$ & $\Delta_y(y)$ & $B(x,y)=0$\\
\hline \hline 1 & $> 0$ & $\geq 2$ & $\leq -2$& $ -\gamma $ &
$\gamma$ &
\begin{picture}(60,60)(-30,-30)
\put(-28.5,-30){\includegraphics*[height=2cm,width=2cm]{case1.bmp}}
\end{picture}
&
\begin{picture}(60,60)(-30,-30)
\put(-28.5,-30){\includegraphics*[height=2cm,width=2cm]{case2.bmp}}
\end{picture}
&
\begin{picture}(60,60)(-30,-30)
\put(-28.5,-30){\includegraphics*[height=2cm,width=2cm]{case1a.bmp}}
\end{picture}\\
\hline 2 & $< 0$ & $\leq -2$ & $\geq 2$& $ -\gamma $ & $\gamma$ &
\begin{picture}(60,60)(-30,-30)
\put(-28.5,-30){\includegraphics*[height=2cm,width=2cm]{case2.bmp}}
\end{picture}
 &
\begin{picture}(60,60)(-30,-30)
\put(-28.5,-30){\includegraphics*[height=2cm,width=2cm]{case1.bmp}}
\end{picture}
 &
\begin{picture}(60,60)(-30,-30)
\put(-28.5,-30){\includegraphics*[height=2cm,width=2cm]{case2a.bmp}}
\end{picture}\\
\hline
\end{tabular}
\caption{The possible forms of $\Delta_x(x)$ and $\Delta_y(y)$,
see (\ref{bdescx}) and (\ref{bdescy}), where $\gamma \neq 0$, and
corresponding representative asymmetric biquadratics $B(x,y)=0$
of (\ref{basym}).}
\end{figure}
The functions $\Delta_x(x)$ and $\Delta_y(y)$ each take two
different possible forms
\begin{eqnarray}
  \gamma\left(-x^4+\frac{\gamma^2+\epsilon^2+1 }{ \gamma}x^2-1\right)\, , \; \;
  \gamma>0, \\
  \gamma\left(y^4+\frac{\gamma^2+\epsilon^2+1 }{ \gamma}y^2+1\right)\, , \; \;
  \gamma>0
\end{eqnarray}
and
\begin{eqnarray}
  -\gamma\left(x^4+\frac{\gamma^2+\epsilon^2+1 }{ -\gamma}x^2+1\right)\, , \; \;
  \gamma<0, \\
  -\gamma\left(-y^4+\frac{\gamma^2+\epsilon^2+1 }{ -\gamma}y^2-1\right)\, , \; \;
  \gamma<0.
\end{eqnarray}
Comparing the shape of $\Delta_x(x)$ and $\Delta_y(y)$ given in
Figure 1 above with those given in \cite[Figure 7]{aijrii}
suggests the following parametrizations:\newline

Case 1. $\mathrm{dn}/\sqrt{k'}$ or $\sqrt{k'}\mathrm{nd}$ for $x$
and $\mathrm{cs}/\sqrt{k'}$ or $\sqrt{k'}\mathrm{sc}$ for $y$, and

Case 2. $\mathrm{cs}/\sqrt{k'}$ or $\sqrt{k'}\mathrm{sc}$ for $x$
and $\mathrm{dn}/\sqrt{k'}$ or $\sqrt{k'}\mathrm{nd}$ for
$y$.\newline

Following the procedure outlined in \cite{aijrii}, we have
obtained the various possible parametrizations given in Table 1.
Note that ${\cal B}_x = -{\cal B}_y = 1/k +k \ge 2$ for Case 1 and
${\cal B}_x = -{\cal B}_y = -1/k - k \le -2$ for Case 2 when $k\in
[0,1]$. As a result the elliptic parametrizations given in Table 1
cover all possible cases. Also note that in Table 1 there are two
possible parametrizations of $x$, $y$ for each subcase. The second
is related to the first by a shift of the argument $u \rightarrow
u+{\bf K}(k)$, where
\begin{eqnarray}
  {\bf K}(k)=\int_0^1 \frac{dt} {\sqrt{1-t^2} \sqrt{1-k^2
  t^2}}\nonumber
\end{eqnarray}
is the complete elliptic integral of the first kind (and equals
one quarter of the period of $\mathrm{sn}(u,k)$ and
$\mathrm{cn}(u,k)$).
\begin{table}\label{t53}
\centerline{
\begin{tabular}{|c||c|c|c|c|} \hline \vphantom{\Big\|}
Case & $x$ & $y$ & $\gamma$ & $\epsilon$ \\
\hline \hline  ${1}$& $\begin{array}{c}\vphantom{\Big\|}
  \sqrt{k'}\, \mathrm{nd}(u) \\ \vphantom{\Big\|}
  \mathrm{dn}(u)/  \sqrt{k'}
\end{array}$ & $\begin{array}{c}\vphantom{\Big\|}
  \mathrm{cs}(\pm u+\eta)/\sqrt{k'} \\ \vphantom{\Big\|}
  -\sqrt{k'}\, \mathrm{sc}(\pm u +\eta)
\end{array}$ & $k'\, \mathrm{nd}^2(\eta)$ &
$ k^2 \, \mathrm{sd}(\eta)\mathrm{cd}(\eta)$ \\
\cline{2-5}  & $\begin{array}{c}\vphantom{\Big\|}
  \sqrt{k'}\, \mathrm{nd}(u) \\ \vphantom{\Big\|}
  \mathrm{dn}(u)/  \sqrt{k'}
\end{array}$ & $\begin{array}{c}\vphantom{\Big\|}
  \sqrt{k'}\, \mathrm{sc}(\pm u+\eta) \\ \vphantom{\Big\|}
  -\mathrm{cs}(\pm u+\eta)/\sqrt{k'}
\end{array}$ & $\mathrm{dn}^2(\eta)/k'$ &
$ k^2 \, \mathrm{sn}(\eta)\mathrm{cn}(\eta)/k'$ \\
\hline $2$ & $\begin{array}{c}\vphantom{\Big\|}
  \sqrt{k'}\, \mathrm{sc}(u) \\ \vphantom{\Big\|}
  -\mathrm{cs}(u)/\sqrt{k'}
\end{array}$ & $\begin{array}{c}\vphantom{\Big\|}
  \pm \mathrm{dn}(u\pm\eta)/\sqrt{k'} \\ \vphantom{\Big\|}
  \pm \sqrt{k'}\, \mathrm{nd}(u\pm\eta)
\end{array}$ & $-k'\, \mathrm{nd}^2(\eta)$
&
$k^2\, \mathrm{sd}(\eta)\mathrm{cd}(\eta)$ \\
\cline{2-5}  & $\begin{array}{c}\vphantom{\Big\|}
  -\mathrm{cs}(u)/\sqrt{k'} \\ \vphantom{\Big\|}
  \sqrt{k'}\, \mathrm{sc}(u)
\end{array}$ & $\begin{array}{c}
  \mp \mathrm{dn}(u\pm\eta)/\sqrt{k'} \\
  \mp \sqrt{k'}\, \mathrm{nd}(u\pm\eta)
\end{array}$ & $-\mathrm{dn}^2(\eta)/k'$&
$k^2\, \mathrm{sn}(\eta)\mathrm{cn}(\eta)/k'$ \\
\hline
\end{tabular}
} \caption{Elliptic parametrization of the cases given in Figure
1 (the explicit dependence of functions on the modulus is
suppressed for convenience).}
\end{table}
\\

\noindent{\bf Remark 1} The asymmetric biquadratic (\ref{basym})
is invariant under $(x,y) \mapsto (-x,-y)$. As a result the
parametrizations given in Table 1 with $-x$ and $-y$ also
parametrize the curve.
\\

\noindent{\bf Remark 2} Using the parametrizations given in Table
1 and (\ref{bmapasym}) (with $\epsilon \rightarrow 2\epsilon$), we
generally obtain $(x',y')= (-p_x(\pm u+2\eta),-p_y(\pm u+3\eta))$
for Case 1 and $(x',y')= (-p_x(u\pm 2\eta),-p_y(u\pm 3\eta))$ for
Case 2, where $p_x$ and $p_y$ are the appropriate elliptic
functions from the Table.

The above parametrizations for $x'$ and $y'$ can be verified using
the following fact: $B(x,y) = B(x',y) = B(x',y')=0$, where
$B(x,y)$ is given by equation (\ref{basym}). Assume $x$ and $y$
satisfy $B(x,y)=0$ and have the parametrizations $p_x$ and $p_y$,
respectively, where $p_x$ and $p_y$ are appropriate elliptic
functions for Case 1 (Case 2), say. The parametrization for $x'$
can be determined using $B(x',y)=0$, i.e.
\begin{eqnarray}\label{basyma}
  B(x',y)&=&x'^2 y^2 +\gamma (x'^2-y^2) +2\epsilon x' y
  -1\nonumber \\
  &=& y^2 x'^2 - \gamma (y^2-x'^2) +2\epsilon y x'
  -1=0.
\end{eqnarray}
The $y$ variable, whose parametrization is known, is treated as
the first variable in Table 1 and $x'$, whose parametrization is
to be determined, is the second variable. Case 2 (Case 1) is now
being considered. Finally, $B(x',y')=0$ is used to determine the
parametrization for $y'$. In this case $x'$, whose parametrization
is known, is treated as the first variable in Table 1 and $y'$,
whose parametrization is to be determined, is the second variable.
Case 1 (Case 2) is now being considered again. Care needs to be
taken to ensure that the same pair of parametrizations are chosen
and that the right signs appear in the parametrizations for $x'$
and $y'$. This can be achieved using $x$, $y$, $x'$ or $y'$ when
required.
\\

\noindent{\bf Remark 3} The parametrizations given in Table 1
allow an action-angle variable description of the dynamics on each
of the asymmetric biquadratics of Figure 1 under their
corresponding asymmetric maps (\ref{bmapasym}). Identifying $(x,y)
\rightarrow (x_n,y_n)$ and $(x',y') \rightarrow
(x_{n+1},y_{n+1})$, $n \in \ZZ$ being discrete time, (\ref{basym})
becomes
\begin{equation} \label{basym1}
     x_n^2 y_n^2+\gamma(x_n^2-y_n^2)+2\epsilon x_n y_n - 1=0,
\end{equation}
whereas (\ref{bmapasym}) (with $\epsilon \rightarrow 2\epsilon$)
becomes
\begin{equation} \label{bmapasym1}
    x_{n+1}=-x_n-\frac{2\epsilon y_n }{ y^2_n+\gamma}\, , \; \;
    y_{n+1}=-y_n-\frac{2\epsilon x_{n+1} }{ x^2_{n+1}-\gamma}.
\end{equation}
Using Remark 2 above, the mapping (\ref{bmapasym1}) has the
solution
\begin{eqnarray} \label{bparaf}
x_n&=&(-1)^n p_x(u_{2n},k)=(-1)^n p_x(2n \,\eta + u_0,k),\nonumber
\\
y_n&=&(-1)^n p_y(u_{2n+1},k)=(-1)^n p_y((2n+1) \,\eta + u_0,k).
\end{eqnarray}
This now allows us to describe the dynamics on each canonical
asymmetric biquadratic curve in terms of the modulus $k$ and
argument $u$ of the elliptic functions:
\begin{eqnarray}
k_{n+1}&=&k_n \nonumber \\ u_{n+1}&=&u_n + \eta. \nonumber
\end{eqnarray}

Finally, recall that (\ref{basym}), equivalently (\ref{basym1}),
is a normal form for many elliptic asymmetric biquadratics. In
this case the original elliptic asymmetric biquadratic
(\ref{biquad}), written in terms of $(X_n,Y_n)$ (after the
identification $(X,Y) \rightarrow (X_n,Y_n)$ and $(X',Y')
\rightarrow (X_{n+1},Y_{n+1})$), can be related to (\ref{basym1})
using an asymmetric modular transformation
\begin{equation}\label{bdiffmod}
  (X_{n},Y_{n})=(f(x_n),g(y_n)),
\end{equation}
where $f$ and $g$ are modular. It now follows from above that
successive points on the original biquadratic can be written:
\begin{eqnarray}\label{bdiffmod2}
(X_{n},Y_{n})&=&(f(x_n),g(y_n))\nonumber
\\ &=& ((f\circ ((-1)^n p_x)) \,(2n \,\eta + u_0,k),(g\circ ((-1)^n p_y)) \,((2n+1)\,\eta +
u_0,k)).\nonumber \\
\end{eqnarray}

\section{Symmetric Case}

Consider the biquadratic
\begin{equation} \label{abscan}
     B(x,y)=x^2 y^2 +\gamma (x^2 + y^2) +2 \epsilon x y + 1 =0.
\end{equation}
Writing $y$ as a function of $x$, we get
\begin{equation} \label{ayasafxpm1}
     y = \frac{- \epsilon x \pm \sqrt{- \gamma x^4 +
     (\epsilon^2 - \gamma^2 - 1) x^2 - \gamma}}{x^2 + \gamma}.
\end{equation}
The quartic under the square root sign can be written as
\begin{equation}
   -\gamma\left(x^4+\frac{\epsilon^2-\gamma^2-1 }{ -\gamma}x^2+1\right)\, , \; \;
   -\gamma>0.
\end{equation}
The two cases of $B(x,y)=0$ which were not parametrized. using
(ratios of) Jacobian elliptic functions in \cite{aijrii}
correspond to $0 < {\cal B} < 2$ (Case 6 of \cite[Figure
7]{aijrii}) and $-2 < {\cal B} < 0$ (Case 4 of \cite[Figure
7]{aijrii} when $-2 < {\cal B} < 0$), where ${\cal B}$ is defined
as
\begin{equation} \label{adefellb}
{\cal B} = \frac{(\epsilon^2 - \gamma^2 - 1)}{\gamma}.
\end{equation}
The two cases can be combined into one, i.e. $-2 < {\cal B} < 2$.
In what follows, we show that this case can be transformed into
one where a parametrization has been found.

Consider the modular transformation\footnote{We note that for all
parametrizable cases given in \cite{aijrii}, the quartic is
factorizable in the form $(ax^2+b)(cx^2+d)$. For the outstanding
(combined) case, however, the quartic takes the form $x^4 + \alpha
x^2+1$, where $-2<\alpha<2$, which is factorizable in the form
$(x^2+\sqrt{2-\alpha}\, x+1)(x^2-\sqrt{2-\alpha}\, x+1)$. Looking
for a modular transformation that produces a quartic with no odd
terms in the quadratics, we obtain $x=(1-\bar{x}) /(1+\bar{x})$.}
$(x,y)=(\frac{1-\bar{x}} {1+\bar{x}},
\frac{1-\bar{y}}{1+\bar{y}})$. Applying this transformation to the
biquadratic (\ref{abscan}) we obtain
\begin{equation} \label{abscan1}
     \bar{B}(\bar{x},\bar{y})=(1+\gamma+\epsilon)(\bar{x}^2 \bar{y}^2+1)
     +(1+\gamma-\epsilon)(\bar{x}^2 + \bar{y}^2) +4(1-\gamma)
     \bar{x} \bar{y}=0.
\end{equation}
Finally, dividing through by the coefficient of
$(\bar{x}^2\bar{y}^2+1)$ we obtain
\begin{equation} \label{abscan2}
     \hat{B}(\hat{x},\hat{y})=\hat{x}^2 \hat{y}^2
     +\left(\frac{1+\gamma-\epsilon}
     {1+\gamma+\epsilon}\right) (\hat{x}^2 + \hat{y}^2)
     +4\left(\frac{1-\gamma}{1+\gamma+\epsilon}\right) \hat{x} \hat{y}
     + 1 =0.
\end{equation}
In doing the division, we note that the coefficient of
$(\bar{x}^2\bar{y}^2+1)$ is necessarily non-zero. This follows
since the coefficient of $(\bar{x}^2\bar{y}^2+1)$ being zero
implies the biquadratic, and the initial curve from which it is
transformed, is rational, as can easily be checked.

Writing $y$ as a function of $x$, we get
\begin{equation} \label{ayasafxpm2}
     y = \frac{2 (\gamma-1) x \pm \sqrt{(\epsilon^2-(\gamma+1)^2)
     (x^4+1)-2(\epsilon^2 - (\gamma+1)^2 +8\gamma) x^2}}
     {(1+\gamma+\epsilon)x^2 + 1+\gamma-\epsilon}.
\end{equation}
The quartic under the square root sign can be rewritten as
\begin{equation}
   (\epsilon^2-(\gamma+1)^2)\left\{x^4
   -\left(2+\frac{16\gamma}
   {\epsilon^2-(\gamma+1)^2}\right)x^2+1\right\}\, , \; \;
   -\gamma>0.
\end{equation}
Note that the term $\epsilon^2-(\gamma+1)^2$ is positive, see
(\ref{adefb1}) below. To determine the range of $-2-16\gamma
/(\epsilon^2-(\gamma+1)^2)$ we can use $-2 < {\cal B} < 2$ from
above, i.e.
\begin{equation} \label{adefb}
-2 < \frac{(\epsilon^2 - \gamma^2 - 1)}{\gamma} < 2.
\end{equation}
Subracting 2 and then dividing by -16 we obtain
\begin{equation} \label{adefb1}
0 < \frac{\epsilon^2 - (\gamma^2 + 1)^2}{-16\gamma} < \frac{1}{4}.
\end{equation}
Inverting and then subtracting 2 we obtain
\begin{equation} \label{adefb2}
2 < -2-\frac{16\gamma}{\epsilon^2 - (\gamma^2 + 1)^2} < +\infty.
\end{equation}
I.e. Case 4 of \cite[Figure 7]{aijrii}) when ${\cal B} < - 2$,
where ${\cal B}$ is defined as ${\cal B}=2+16\gamma
/(\epsilon^2-(\gamma+1)^2)$, which is parametrized by $\sqrt{k'}\,
\mathrm{sc}$ or $\mathrm{cs}/\sqrt{k'}$, see \cite[Table
3]{aijrii}).

\section*{Acknowledgements}

Part of this research was undertaken while in receipt of a La
Trobe University Postgraduate Research Scholarship.

\end{document}